# Vortex rectification effects in plain superconducting films

Farkhad G. Aliev [1*], Arkady P. Levanyuk[1], Raúl Villar[1], Elvira M. Gonzalez[2] and Victor V. Moshchalkov[3]

(1) Instituto Nicolás Cabrera de Ciencia de Materiales, Dpto Física de la Materia Condensada, C-III, Universidad Autónoma de Madrid, 28049, Madrid, Spain

(2) Departamento de Física de Materiales, Universidad Complutense de Madrid, 28040, Madrid, Spain

(3) Nanoscale Superconductivity and Magnetism Group, LVSM, Katholieke Universiteit Leuven, Celestijnenlaan 200 D, B-3001 Leuven, Belgium

(*) Corresponding author
E mail: farkhad.aliev@uam.es

Superconducting films in perpendicular magnetic field are found to rectify alternating currents. The effect has been observed both in plain and nanostructured superconducting films (niobium and lead). The rectified voltage appears both along and transverse to alternating current and strongly depends on the magnetic field. Rectification phenomena is based on the property of superconductors to have permanent circulating currents in magnetic field (vortices) and to develop electric fields at high enough currents. In contrast to previous findings, at higher current frequencies no asymmetric pinning sites are needed to produce rectification and related guided flux motion.

The main idea for controlled vortex motion in a superconductor is based on the application of a well defined drive (Lorentz force), which can be provided by DC current sent through the superconductor in presence of an applied magnetic field. Additionally, the pinning landscape for vortices can be also tuned by using nano-engineered artificial pinning centres and AC instead of DC current. Recently several realizations have been proposed where guided vortex motion can be achieved by fabricating asymmetric pinning sites ("vortex ratchets") (1-5). Villegas *et al.* (6) presented experimental data on superconducting films with patterned periodic arrays of triangular magnetic dots. They found that when an AC current is applied to the film in a transversal magnetic field a DC voltage appears in the direction of the current. In other words, there is a permanent vortex flow perpendicular to the current generating the measured DC voltage between the potential probes. This finding was interpreted as a manifestation of the above mentioned ratchet effect.

We focus here on the investigation of an even more pronounced rectification effect in plain superconducting thin films *without any prefabricated pinning centres* and also in films with prefabricated periodic arrays of *symmetrical* pinning centres (cylindrical antidots). The difference between plain and nanostructured films is that the introduced periodic pinning centers (PPCs) may be additionally used to effectively control sign and value of the rectification voltage. Interestingly, in nanostructured superconductors rectification can be inversed not only by inverting the external magnetic field but also by tuning the number of vortices per pinning centre. This new form of AC current rectification in superconductors is related to guided one-dimensional flow of vortices ("vortex rivers") which could be used for studies of vortex pinning phenomena in a broad temperature range below the superconducting critical temperature.

The experiments were performed on plain and nanostructured Nb and on nanostructured Pb thin films optically patterned to "cross" configuration (cross area varied between 40*40 and 300*300 $\mu m^2$), which allowed using four current and four voltage probes (see upper inset to Figure 1). A cross-shaped configuration has recently been used (5,7) to study transport properties for different orientations of current and electric field. Besides plain superconductor Nb films with thickness of 100nm, we have studied nanostructured Pb and Nb films with different pinning centers symmetry. Namely, Pb film (thickness of 50nm covered with 20 nm of Ge, structured with square periodic pinning array of circular holes with diameter of 0.6 nm and period 1.5 $\mu m$) and Nb film (with 100 nm thick with rectangular array of PPCs (0.4*0.5 $\mu m^2$) of circular Ni dots with 250 nm diameter). Further information about films growth and characterization can be found in references (7-8). Electron transport was studied with and without sinusoidal AC drive current of frequency in the range $10^4$ - $1.5*10^8$ Hz, which was injected through oppositely situated current contacts. Simultaneously a DC voltage was measured by a nanovoltmeter longitudinally to the current (between oppositely situated contacts $U_{1-2}$ or $U_{3-4}$) or in the transverse geometry ($U_{2-3}$ or $U_{4-1}$) perpendicular to the AC drive current. The constant magnetic field was perpendicular to the film surface. A two-loop temperature control provided temperature stability better than 0.2 mK during measurements.

In Figure 1 (upper right inset) we compare the DC resistance measured through the normal-superconductor transition (right axis) with the DC voltage between the same contacts for the same set of external parameters, AC current, frequency and magnetic field, except that the DC current is kept zero. No finite DC voltage appears above $T_c$ in any geometry. The rectified signal is observed far below $T_c$, showing a strongly non monotonous highly reproducible temperature dependence with several sign inversions. The value and polarity of the DC voltage generated in the superconducting state in presence of AC drive is mainly controlled by the applied magnetic field (independently of the magnetic history), AC drive intensity and, more weakly, by the drive frequency (see Figure 2). In this paper we present data obtained for frequencies above $10^7$ Hz, where rectification effects are most pronounced. Despite the complexity of $U_{DC}(T)$, the main features (i.e. the two minima marked by arrows in Figure 1) scale with AC current intensity and are shifted in temperature. Left bottom inset to Figure 1 shows the DC voltage dependence of the minimum marked by the red arrow on AC current, which is clearly non-linear.

In Figure 3 the temperature dependences of the rectified voltage, measured with the same AC drive from four different pairs of contacts moving " counter clock wise", are compared. Note that close to $T_C$ the rectified voltage has nearly the same absolute values and different signs for opposite sides and opposite magnetic fields. The situation is different at temperatures $T < 0.8T_c$, where rectification is dominant on neighbouring sides of the cross. In all cases the sum of all four potential differences is close to zero, which means that the rectified potential difference round the path integral along the cross border is nearly zero: $\int \mathbf{E}d\mathbf{l} \approx 0$ (here $\mathbf{E}$ is electric field). Therefore, the AC drive does not contribute to the mean value of the magnetic flux inside the cross area. The rectification effects that we observe are clearly related to vortex pinning regimes, and they disappear below the temperatures at which the vortices become strongly pinned.

We have also found rectification in nanostructured Pb and Nb films with *symmetric* pinning centers, where pinning is stronger and obviously better controlled than in plain films (8,9). However, the AC current rectification in superconductors with PPCs reveals at least three main differences in comparison with plain films. First of all, much stronger vortex pinning in superconductors with PPCs increases the critical current $J_C$ (9) and lowers (in about an order of magnitude) the DC voltage. Secondly, the rectified voltage is almost independent of magnetic field polarity in the nearest proximity of the superconducting transition. Finally, and most interestingly, beyond this interval the DC voltage reverses not only with the direction of the magnetic field (as in plain films) but also when the intensity of the magnetic field crosses the matching fields.

In Figure 4 we present data only for Pb film with PPCs, where the oscillation of DC voltage with magnetic field is found to be more pronounced. Upper inset in Figure 4 shows that in Pb with PPCs low temperature contribution to the rectified voltage is apparently non monotonous up to higher matching fields $H_n$ showing oscillating behaviour at least up to n=3. It seems interesting to note that the rectified voltage inverts sign not only when crossing integer matching fields but also near rational field $H_{1/2}$ (i.e. near the magnetic field corresponding to on average half of flux quanta per antidot). We conclude that the rectified voltage changes sign when moving from regime with dominating "excess of unpinned vortices" for H< $H_{1/2}$ to the regime with dominating " deficit of unpinned vortices". Details on experimental observations of AC current rectification for Nb films with symmetric PPCs will be subject of a separate forthcoming publication.

Previously, rectification effects in superconductors have been observed *for alternating magnetic fields* (10,11) but surprisingly, to the best of our knowledge, *nobody has reported on AC current rectification effects in superconductors*. In looking for a qualitative explanation of our data we have used some of the ideas of previous authors (12). Let us consider a strip (lower inset in Figure 4). If a transversal magnetic field is applied in the superconductor phase there is a screening current in the strip with the current density equal to its critical value near the lateral boundaries of the strip (13) and with opposite directions at the two boundaries. Apply now an external current along the strip. The current distribution in the strip will depend on the character of the total current increase but, anyway, one can expect that close to side of the strip the current is lower than its critical while close to the other it is higher. Then the electric field appears only close to one side of the strip [cf. (12)]. If the applied current is of opposite direction the electric field, of opposite direction, appears close to the other side of the strip. If the change of the total current is stopped the electric field will relax, of course, but during some time a relatively high voltage can be observed in a finite strip. In a finite strip the voltage will be both longitudinal and transversal while in an infinite strip only a transversal voltage is expected to survive. For $T=0.93T_c$ and $H=100G$ we estimate that the local excess of the current density over the critical one ($J_c$) should be about (0.5-1) $J_c$ to obtain the experimentally observed voltage. Gurevich *et al*. (14) have argued that the duration of the first stage of relaxation of vortices is a macroscopic quantity which can be expressed via directly measured parameters. Despite they considered a different geometry (slab of width *a* in parallel magnetic field and relaxation after the magnetic field change), nevertheless, their estimation of the characteristic time: $\tau \approx \mu_0(\partial j/\partial E)a^2$ can be considered as a result of a dimensional analysis and could be applied for rough estimation to our case. If we take $\partial j/\partial E$ of about its value in the normal phase and $a^2$ about cross area, we get $\tau \sim 10^{-7} - 10^{-8}$ sec . This is in reasonable agreement with the frequencies for which the effect becomes clearly observable. In the case considered in ref. (15) $\tau$ was proportional to the characteristic time of change of the external magnetic field. Similar behaviour in our case could explain the weak frequency dependence of the rectification effect.

The above arguments imply that the sign of the voltage between the same lateral contact should change the sign if the direction of permanent field is changed. This is, indeed, the case (see Fig.1). Another natural conclusion is that the electric fields at the opposite lateral sides are of the same value and of opposite signs. This is not exactly the case experimentally. Near $T_c$ the signs are, indeed, opposite but the absolute values are not equal (Figure 3). Quite spectacular is the change of the value and even of the sign of the effect with change of temperature. At the moment we have no explanation of this phenomenon. Finally we mention that in the case with superconductors with PPCs ramping of the field through matching fields also changes sign of rectified voltage. This suggests that the dependence of the critical current on the field is important for the phenomena because it is just the sign of derivative $dJ_c/dH$ which changes when crossing a matching field (15).

We thank V.Pryadun, R.Guerrero, M.Alieva, T.Alieva, J.Sierra and D.Golubovic for assistance and G.Mikitik, J.J.Palacios, J.L.Vicent, S.Vieira for discussion.

Correspondence should be addressed to F.G.A. (e-mail: farkhad.aliev@uam.es)

Figure Captions

Figure 1. Rectified DC voltage as a function of temperature in Nb film measured in transverse geometry ($U_{2-3}$). The different curves show transverse DC voltage measured for different AC drives with frequency f = 43 MHz, and a magnetic field H = 100G in a wide temperature range below $T_c$. For $I_{AC}$ = 2.1 mA the data were obtained both decreasing and increasing the temperature. Note that similar temperature dependences are obtained for different AC drives, but they are shifted in temperature. This shift is not related to any heating effect because it is at least one order of magnitude larger than the reduction of $T_C$ induced by the AC current. This reduction of $T_C$ may be observed at the right end of the curves where the voltage turns to zero. Upper right inset: data for two opposite magnetic fields (±100G) with $I_{AC}$ = 3.6mA and at a higher frequency f = 147 MHz. For comparison we present the temperature dependence of the resistance (superconducting transition) measured between U2 and U3 with a longitudinal DC current I=1 mA in the presence of the same AC current and with a magnetic field of 100 G. Left bottom inset: rectified voltage dependence on $I_{AC}$ of the minimum in $U_{DC}(T)$ marked by the red arrow. Upper left inset shows a typical sample contact configuration and numbers the four measurement sites.

Figure 2. Dependence of DC voltage generated in the superconducting state in presence of AC current on the drive frequency. Figure demonstrates longitudinal ($U_{3-4}$) AC current rectification effect measured in 40*40 µm² Nb cross-shape film for frequencies

from 9.9kHz to 147 MHz for the same AC drive current (1.2mA) and same applied magnetic field (-100G).

Figure 3. Rectified potential difference in Nb film measured around the cross area ( $U_{1-2}$, $U_{2-3}$, $U_{3-4}$ and $U_{4-1}$). In upper inset the longitudinal rectification effect measured from opposite cross sides is compared inverting the magnetic fields. The data both in the main part and in the upper inset were measured with f=43MHz, $I_{AC}$=1.2mA and H=100G. The electric potential profiles around the cross are schematically sketched near $T_C$ (lower right inset) and far below $T_C$ (lower left inset). The appearance of a non-zero electric field in type II superconductors below $T_C$ indicates a finite average flux of vortices between the corresponding points. This means that far below $T_C$ the movement of vortices has predominantly 1D character as indicated by the arrow on the bottom left inset while closer to $T_C$ our AC drive pump gives rise to a more symmetric vortex flow: vortices enter the cross area from a pair of opposite sides and leave from the other pair of sides, keeping constant on average the total number of vortices within the cross area. Dominant vortex flow is shown by arrows.

Figure 4. Rectified voltage $U_{2-3}$ as function of temperature in Pb film with periodic pinning centres. The data have been obtained with magnetic fields close to the first matching field $H_1$ = 9.2G, and with f=147MHz, $I_{AC}$=3.4mA. Close to $T_C$ the DC voltage is an antisymmetric function of the magnetic field (not shown) while far below $T_c$ it changes sign both with reversion of the magnetic field and when its intensity crosses the matching fields $H_n$=n×$H_1$, which are integer multiples of $H_1$. The latter corresponds to one flux quantum per antidot. The upper inset shows the dependence of the amplitude of the low temperature anomaly $U_A$ near 5.5K, signalled by an arrow, as a function of the magnetic field normalized by the first matching field $H_1$, measured with the same frequency and drive current. The antisymmetry of this low temperature anomaly in respect to $H_n$ gradually smears out for n>3 (not shown). The lower inset is an illustration to the text.

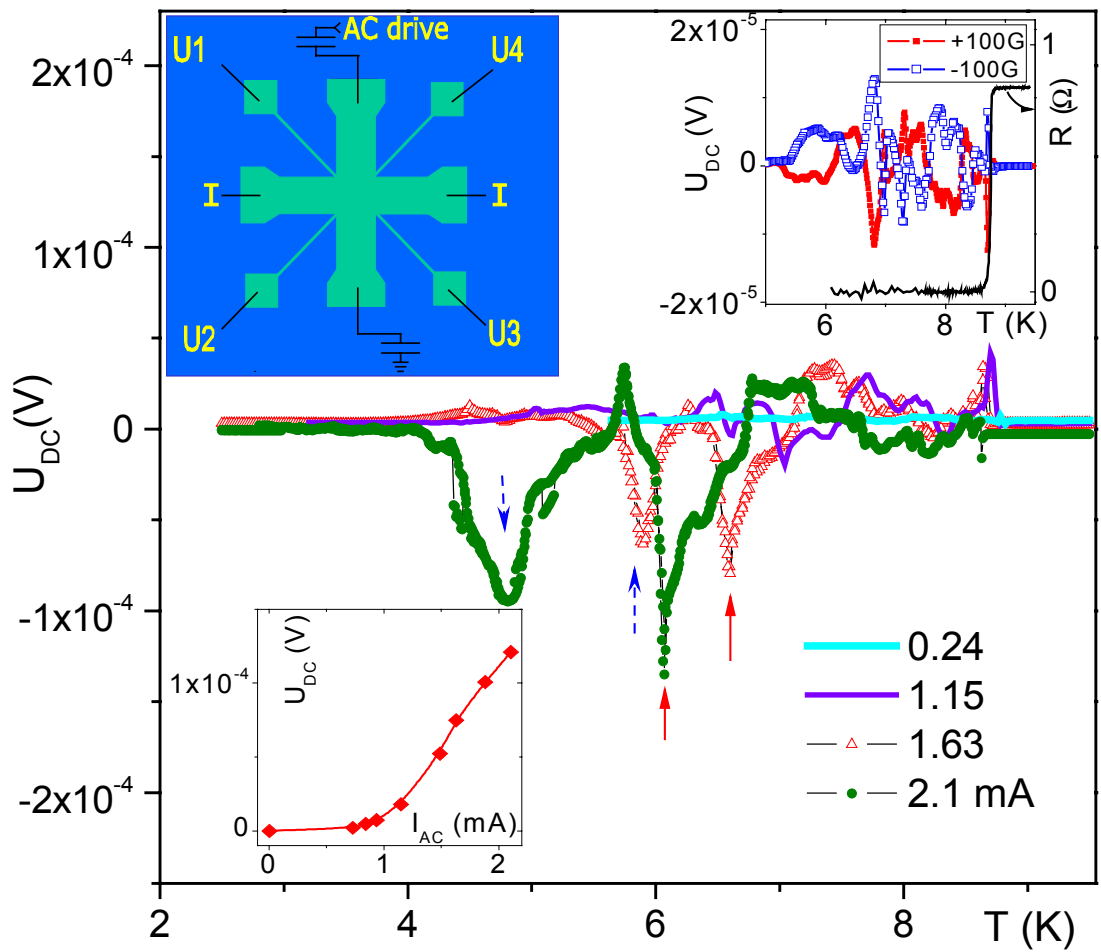

Figure 1

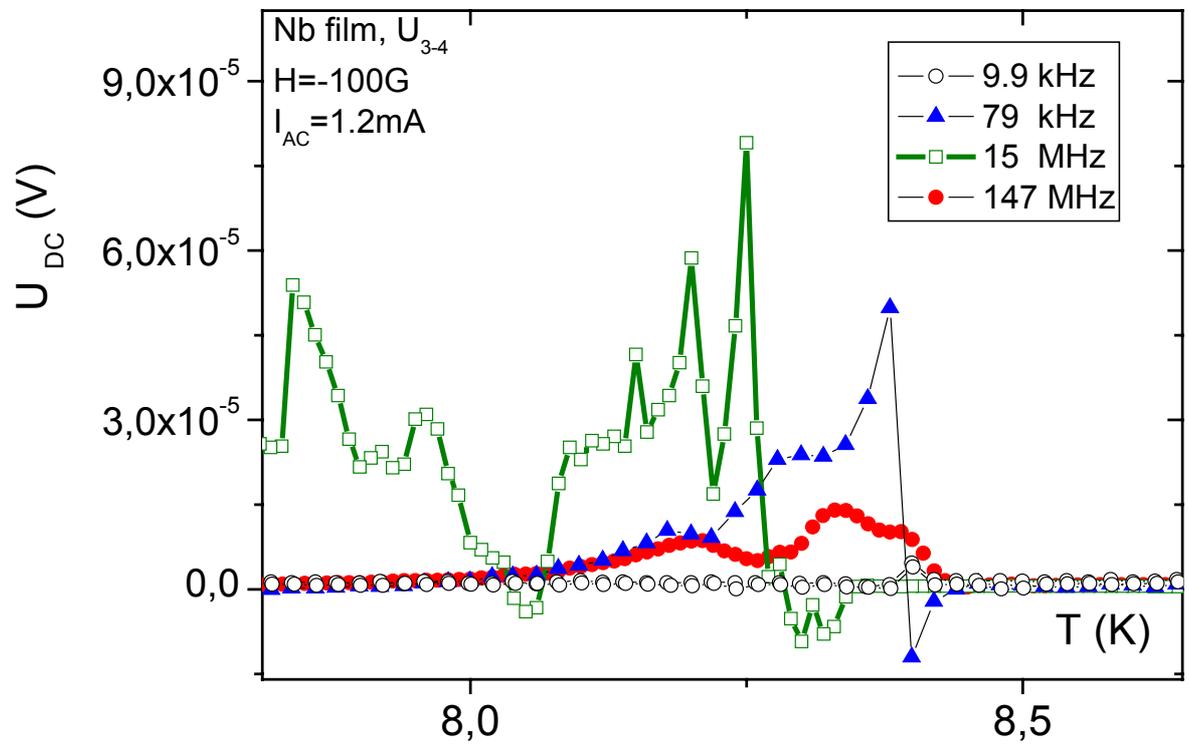

Figure 2

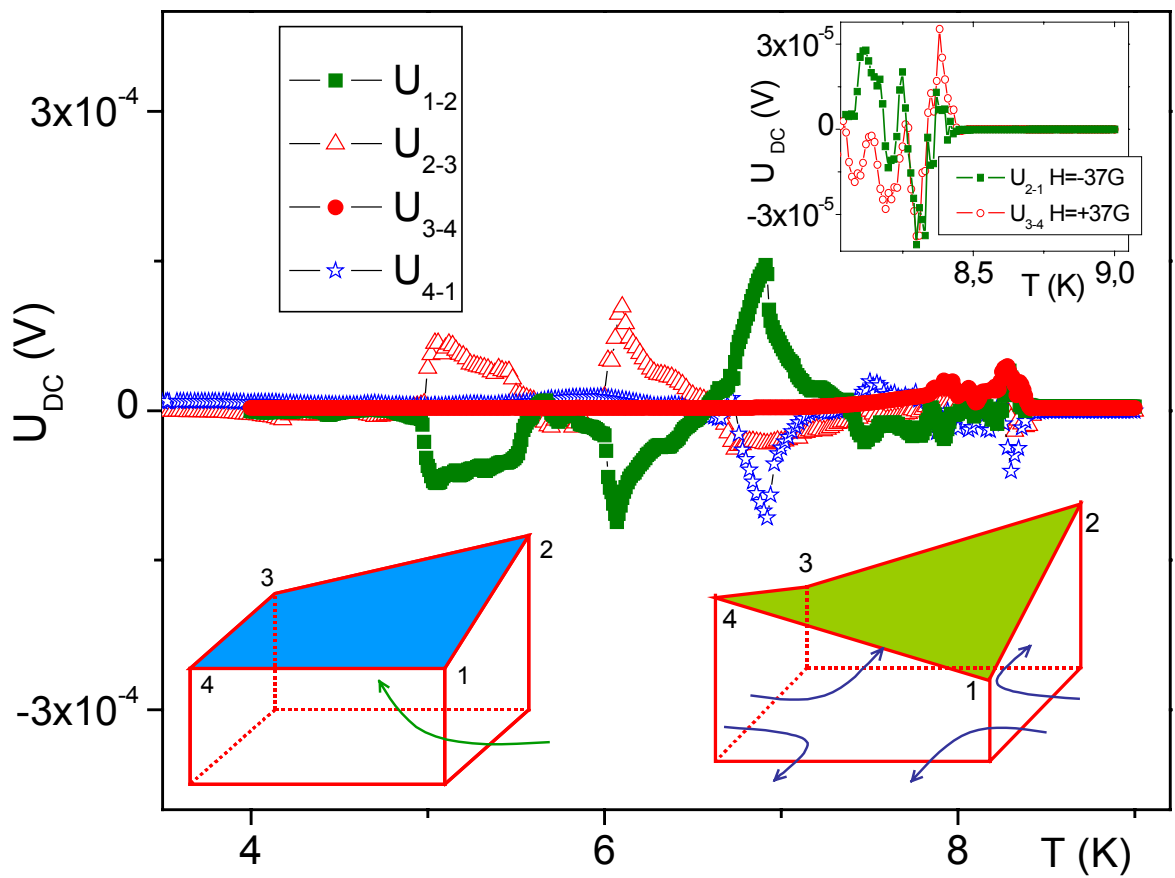

Figure 3

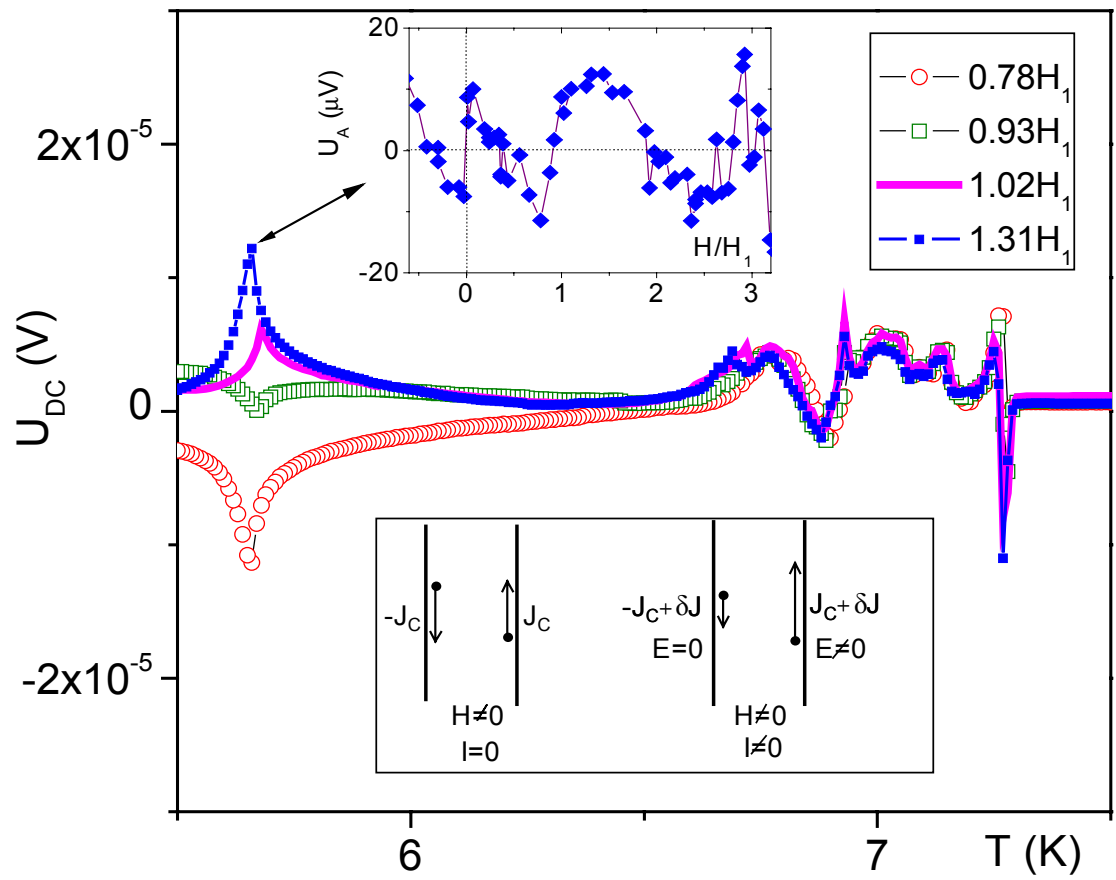

Figure 4